\documentclass[final]{appolb}

 \usepackage{epsfig}
 \usepackage{amsmath}

\def\beq{\begin{equation}}
\def\eeq{\end{equation}}

\def\tw{\textwidth}

\def\lrang#1{\left\langle#1\right\rangle}

\def\eV{{\rm e\kern-0.12em V}}            
 \def\GeV{{\rm G}\eV} 
\def\cO#1{{\cal O}\left(#1\right)}

\def\cV{{\cal V}}
\def\asPT{\alpha_{\mathsf{s}}^\mathsf{PT}}

\newcommand{\as}{\alpha_\mathsf{s}}
\newcommand{\CF}{C_F}

\def\LQCD{\Lambda_{\mbox{\scriptsize QCD}}}

\def\la{\mathrel{\mathpalette\fun <}}
\def\ga{\mathrel{\mathpalette\fun >}}
\def\fun#1#2{\lower3.6pt\vbox{\baselineskip0pt\lineskip.9pt
  \ialign{$\mathsurround=0pt#1\hfil##\hfil$\crcr#2\crcr\sim\crcr}}}

\begin{document}
\pagestyle{plain}
%% uncomment the following line to get equations numbered by (sec.num)
%\eqsec
%\newcount\eLiNe\eLiNe=\inputlineno\advance\eLiNe by -1

\title{Historical and futuristic perturbative 
       and non-perturbative aspects of QCD jet physics}%
\author{Yuri Dokshitzer
\address{LPTHE, University Paris-VI, 4
 place Jussieu, F-75252 Paris, France\\ and\\PNPI, 188350 Gatchina,
 St.\ Petersburg, Russia}}
\maketitle

\begin{abstract}
  A brief review of jet physics is presented with an emphasis upon
  open theoretical problems (non-perturbative domain; hadronization
  and confinement) and new phenomena (hadroproduction in heavy ion
  collisions).
\end{abstract}

\section{Prehistory}

 Thirty years ago was the epoch of ISR and SPEAR --- the first
 Jet labs.
 In the realm of {\em hard interactions}, R.~Feynman invented his
 famous ``plateau'' --- {$\ln E$} hadrons streaming from a single
 quark-parton that is struck out of the target proton in a Deep
 Inelastic Scattering process.  Moving from the high energy side ({\em
 soft interactions}\/) V.~Gribov, motivated by the Pomeron picture,
 drew an energetic hadron fluctuating into {$\ln E$} partons. Already
 then the key word {{\em duality}}\/ was pronounced in the context of
 the inter-connection between {partons} and hadrons.

 Those, however, were the times of {\em pictures}, of {\em physical
 intuition}, rather than {\em theoretical expectations}, let alone
 {\em predictions}.  So that as late as 1976, three-jet pioneers still
 spoke about logarithmic multiplicity as being merely ``{\em
 fashionable}''.

 The notion of hadron jets goes back to the dark times of the
 exclusive dominance of cosmic ray physics. On the theory side, the
 existence of jets was envisaged from ``parton models'' in the early
 70's.
% by S.D.Drell, D.J.Levy \&\ T.-M.Yan (1969-70)
%    and N.Cabibbo, G.Parisi \&\ M.Testa (1970).
 In 1974 J.B.~Kogut and L.~Susskind 
% (who were playing with scale invariant parton model) 
 have remarked that hard gluon bremsstrahlung off the {$q\bar{q}$}
 pair {\em may be expected}\/ to give rise to three-jet events in the
 {$e^+e^-$} annihilation into hadrons.

 The first thorough analysis of three-jet events in the QCD context
 was due to John Ellis, Mary K.Gaillard \&\ Graham G.Ross (1976).  It
 is instructive, quoting~\cite{EGR}, to recall the background
 knowledge the authors of this seminal study relied upon:
\begin{itemize}
 \item ``{\em no direct experimental evidence yet exists for gluons}\/''
 (except possibly the fact that not all the nucleon's momentum is
 carried by known quark constituents); 
\item ``{\em there is no direct
 evidence for asymptotic freedom}\/'' (though there may be some deviations
 from scaling in DIS at high $Q^2$)
 \item ``{\em fashion sets}\/'' {$\>\>
 \alpha_V(Q^2)\sim 0.2$ -- $1\>$}  at {$ Q^2\sim 10\,\GeV^2$}. 
\end{itemize}
 It is amazing how far this brave youth managed to leap forward from
 such a shaky base!  They professed {\em coplanar structure}\/ of
 final states and cross section {\em scaling}\/ in $x_T=2p_T/Q$, and,
 having studied vector- versus scalar-gluon cases, managed to rightly
 guess the $\sim$ 10\%\ rate for three-jet events.

 EGR also verified the asymptotic 2-jetness of {$e^+e^-$} annihilation
 events that had been advocated one year earlier by George Sterman
%~\cite{Sterman1975} 
 and concluded
\begin{quote}
 {\em Present ideas about quark/gluon {metamorphosis} into hadrons
 suggest a third jet should exist in the direction of the large
 $\>p_T\>$ particle.} [\ldots]
 {\em This {prejudice} is comforted by the observed jet structure in
 large {$\>p_T\>$ $pp$} collisions.}
\end{quote}
Moreover, they drew a picture with two hadron chains stemming
from gluon fragmentation and remarked, without much ado,
\begin{quote}
 {\em Looking at {\em [this picture]} one might {naively expect} more
 hadrons to be produced in {gluon} fragmentation than in {quark}
 fragmentation, and therefore that {$f(x)$} for gluons should be more
 concentrated at low {$x$}}.
\end{quote}

 \section{Colour in hadroproduction and QCD jets}

 That {\em naivete} was to become the core of the colour driven
 picture of parton hadronization elaborated by Bo Andersson, G\"osta
 Gustafson \&\ Carsten Peterson in 1977.

 Have we learned anything new about the {EGR} {\em
 ``metamorphosis''}\/? Not much, to be honest. At the qualitative
 level we keep following {\em ``the fashion''}\/ --- the classical
 Kogut--Susskind ``vacuum breaking picture''.

\noindent
\begin{minipage}{0.5\tw}
 In a deep inelastic scattering process a {\em green}\/ quark in the
 proton is hit by a virtual photon. The quark leaves the stage and the
 Colour Field starts to build up. A {\em green--anti-green}\/ quark
 pair pops up from the vacuum, splitting the system into two {\em
 globally blanched} sub-systems.
\end{minipage} 
\hfill
\begin{minipage}{0.45\tw}
 \epsfig{file=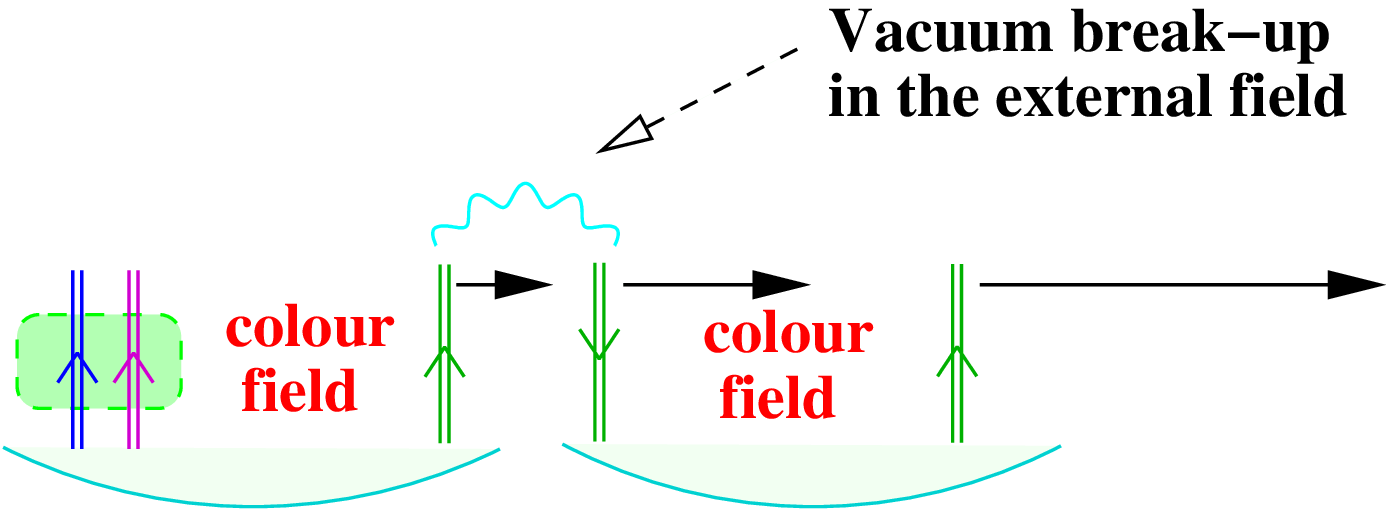,width=0.95\tw,height=0.6\tw}
\end{minipage}

 The Kogut--Susskind scenario had been realized by the
 phenomenological Lund model of multiple hadroproduction \cite{Lund}.
 The Lund hadronization model embodied the key features of the
 Kogut--Susskind scenario namely, the uniformity in {rapidity},
 $\displaystyle {{dN_h}} \propto {{d\omega_h}/{\omega_h}} =
 d\Theta_h/\Theta_h$, and limited transverse momenta of produced
 hadrons with respect to the jet direction.
 
 Picturing a jet as a string of hadrons, the Lund model however made
 an essential step beyond the naive parton model picture by putting a
 special stress on the {\em r\^ole of colour}\/ in hadronization of
 parton ensembles. This has brought to life {\em radiophysics}\/ of  
 QCD jets.

 Let me remind you that studying both {Inter-Jet} and {Intra-Jet}
 phenomena fully revealed colour coherence in QCD parton
 multiplication~\cite{KO}.  Their solid imprint upon the {\em
 angular}\/ and {\em energy}\/ spectra of relatively soft hadrons has
 sent us a powerful message namely, that the confinement (={\em
 metamorphosis}\/) is {\em soft}.

 This is a free lunch that we have not yet found enzymes to digest.
 For the time being, we are {\em exploiting}\/ this gift: Hadron flow
 practitioners (who are developing smart tools for triggering on new
 physics), {Colour Glass} brewers, small-$x$ BFKL lovers, --- no-one
 would hesitate to put {\em gluons}\/ and {\em hadrons}\/ into
 one-to-one correspondence as soon as final state particle production
 issues come onto the stage.

 There is nothing wrong with this. In so doing we simply follow the
 opportunists' motto ``{\em ain't broken -- don't fix it}\/''.  It
 becomes mandatory, however, that we start {\em exploring}\/ the LPHD
 gift rather than simply {\em exploiting}\/ it.  To set up the Quest,
 we have to turn now to the problematics of the {\em
 non-perturbative}\/ domain: what is it, {\em what do we know}\/ about
 it, and, more important still, {\em what we don't}.

 \section{Non-perturbative effects in Event Shape observables}

 There is a specific {(though not too narrow a)} class of QCD
 observables that taught us a thing or two about {genuine
 non-perturbative effects} in multiple production of hadrons in hard
 processes. Among them --- the so-called {\em event shapes}\/ which
 measure global properties of final states {(jet profiles)} in an
 inclusive manner.

 In {$e^+e^-$}, for example, one defines

\begin{minipage}[t]{0.5\tw}
{\begin{align}
\mbox{thrust}\quad T &= \max_{\vec n} \frac{ \sum_i |\vec p_i . \vec
  n|}{\sum_i |{\vec p}_i|}\,, \nonumber\\
\mbox{$C$-param.}\quad C &=  \frac32 \frac{\sum_{i,j} |{\vec p}_i|
  |{\vec p}_j| \sin^2 \theta_{ij}}{\left(\sum_i |{\vec
  p}_i|\right)^2}\,,\nonumber \end{align} }
\end{minipage}%
\hfill
\begin{minipage}[t]{0.4\tw}
\vspace{2mm}
  \resizebox{!}{0.40\textwidth}{\begin{picture}(0,0)%
\includegraphics{tube+labels.pstex}%
\end{picture}%
\setlength{\unitlength}{4144sp}%
\begingroup\makeatletter\ifx\SetFigFont\undefined%
\gdef\SetFigFont#1#2#3#4#5{%
  \reset@font\fontsize{#1}{#2pt}%
  \fontfamily{#3}\fontseries{#4}\fontshape{#5}%
  \selectfont}%
\fi\endgroup%
\begin{picture}(5614,2049)(879,-1873)
\put(6391,-1276){\makebox(0,0)[rb]{\smash{\SetFigFont{17}{20.4}{\familydefault}{\mddefault}{\updefault}
${\vec n}_T$
}}}
\end{picture}
}
\end{minipage}

\[%
\mbox{jet-mass}\quad\,\, \rho =
 \frac{\left(\sum_{i\in\mathrm{hemisphere}}\; p_i \right)^2}{
 \left(\sum_i E_i\right)^2}\,, \qquad\
\mbox{broadening}\quad\!\! B_T = \frac{ \sum_i p_{ti}}{\sum_i
    |{\vec p}_i|}\,. \]

\noindent
 (The two hemispheres in the jet-mass and transverse momentum
 component in broadening are defined with respect to the thrust axis.)
 These and similar event shape observables are formally calculable in
 pQCD (being collinear and infrared safe, CIS) but possess large
 non-perturbative {$1/Q$}--{suppressed corrections}. Being
 perturbatively calculable does not imply, however, being insensitive
 to non-perturbative physics~\cite{NP}.

\noindent
\begin{minipage}[t]{0.49\tw}
        \includegraphics[height=0.85\tw,width=\tw]{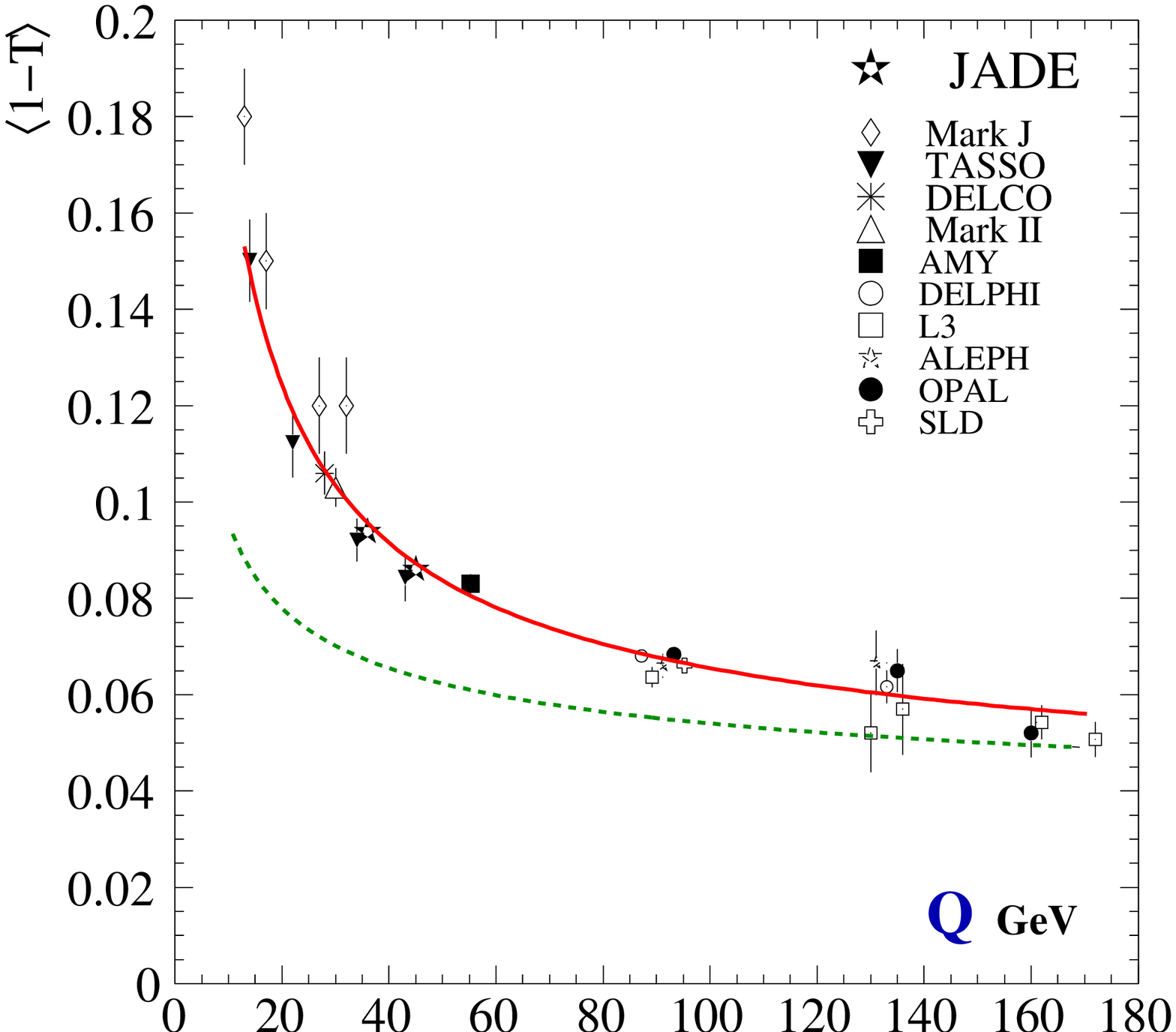}%
        \vspace{-0.2\tw}\\ {\mbox{{ }\qquad\quad Mean Thrust}}%
\end{minipage}%
\hfill
\begin{minipage}[t]{0.49\tw}%
        \includegraphics[height=0.85\tw,width=\tw]{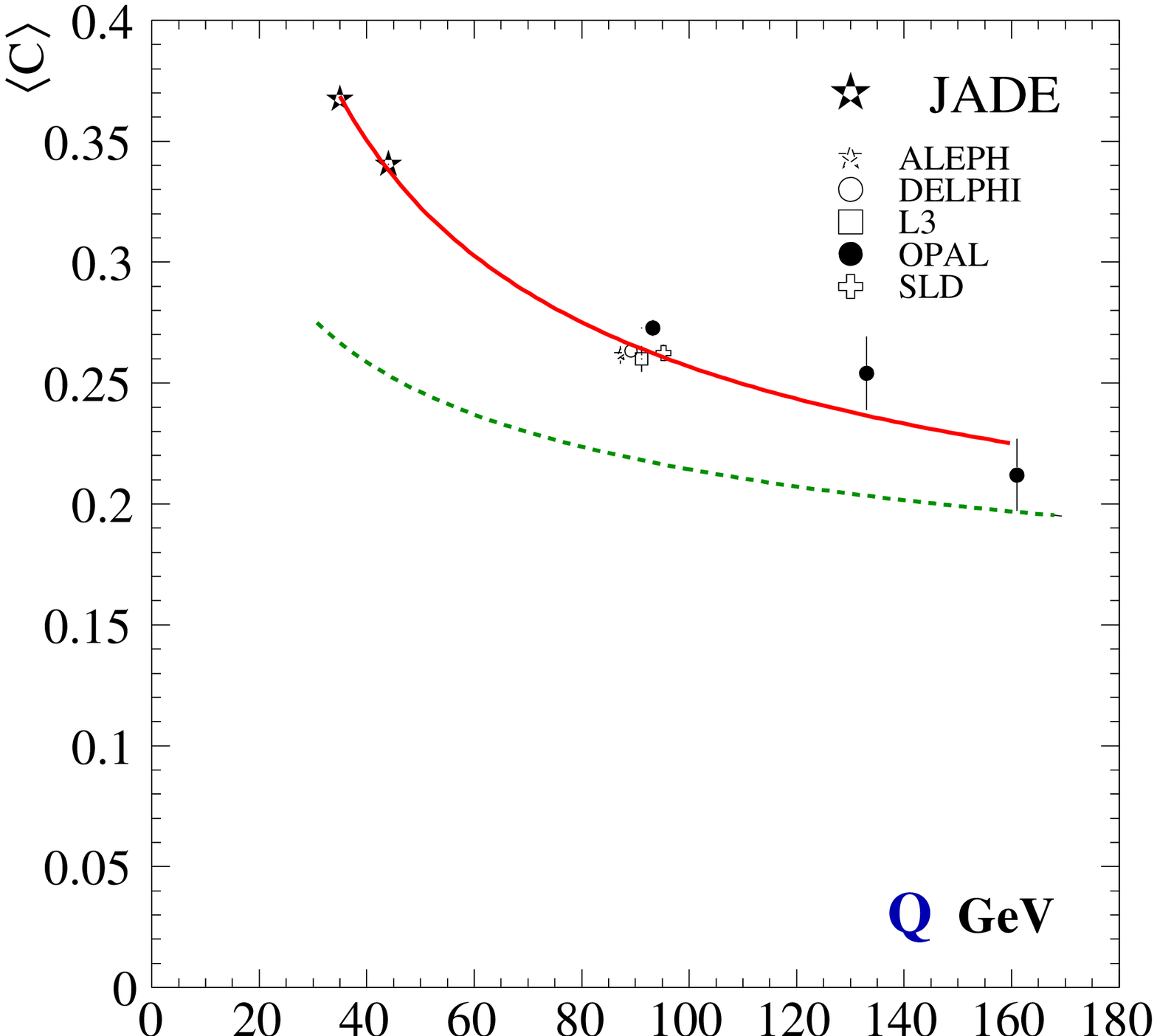}%
        \vspace{-0.2\tw}\\ {\mbox{{ }\qquad\quad Mean
        $C$-parameter}}% 
\end{minipage}

\noindent
 Indeed to reconcile with the data the two-loop pQCD predictions for
 the means of two exemplary jet shapes (shown by dotted lines) one has
 to introduce, on the phenomenology side,
\begin{eqnarray*} 
\lrang{1\!-\!T}_{\mbox{\scriptsize hadron}} \!\!&\approx&\!\! 
\lrang{1\!-\!T}_{\mbox{\scriptsize parton}} \> \>+\>\> {{ 1\>
\GeV}}/{{ Q}}, \\
\lrang{C}_{\mbox{\scriptsize hadron}} \!\!&\approx&\!\! \quad
\lrang{C}_{\mbox{\scriptsize parton}} \quad \!+\> {{4\>
\GeV}}/{{ Q}}.
\end{eqnarray*}

\subsection{NP games}

 The pQCD motivated ``theory'' of genuine non-perturbative effects in
 jet shapes (about 8 years old and running) predicts the above {\em
 ratio}\/ to be
$$ 
%\frac
{3\pi}/{2} \quad \simeq \quad {4}/{{ 1}}
\,. $$

 The origin of power-suppressed corrections to CIS observables can be
 linked with the mathematical property of {badly convergent}
 perturbative series, {typical for field theories,} known under the
 name of ``{renormalons}''.  The renormalon-based analysis is
 perfectly capable of controlling the {\em ratios}\/ of power terms,
 mentioned above, but can say next to nothing about the {\em absolute
 magnitude}\/ of such a correction.  To address the latter issue, an
 additional hypothesis had to be invoked namely, that of the existence
 of an {InfraRed-finite QCD coupling} {(whatever this might mean; a
 detailed discussion of the issue can be found in \cite{Vanc})}.

 \subsection{NP power corrections and IR-finite $\as$}

 The ``{industry-standard}'' {way of fitting event-shape power
 corrections}~\cite{DMW} exploits the idea that the power correction
 is driven by the NP modification of the QCD coupling in the
 InfraRed. The leading NP contribution to a given observable ${\cV}$
 can be parametrized as
\beq\label{eq:NPpar}
    \delta{{\cV}}{_{p}} = { \frac{2\CF}{\pi} \int^{\mu_I}_0
    \frac{dm}{m}\cdot {\biggl(} \frac{m}{Q} {\biggl)^p} \cdot \left(
    \as(m^2) {{- \asPT(m^2)}} \right) } \cdot {{c_\cV}}.
\eeq
The key features of this expression are as follows. 
\begin{itemize}
 \item It contains a PT controlled observable dependent coefficient
 {{$c_\cV$}}; \item The (integer) exponent $p$ determines sensitivity
 of a given observable to the IR momentum scales; for the vast
 majority of event shapes one has {\em linear}\/ damping, ${p =1}$;
 \item One subtracts off {$\asPT$}, the fixed-order perturbative
 expansion of the coupling in order to avoid {double counting}.
\end{itemize}
The full answer for the case under interest, $p=1$, takes the form
\beq\label{eq:fullNP}
  \cV = A \as + B \as^2 + {c_\cV } \,{ \frac{2\CF}{\pi}\frac{\mu_I}{Q}
  \left(\alpha_0 - \langle\asPT\rangle_{\mu_I}\right) }
\eeq
with $\alpha_0$ the {\em first moment}\/ of the coupling in the
InfraRed:
 $$
 \displaystyle {{\alpha_0}} = \frac{1}{\mu_I} \int_0^{\mu_I} dm\,
 \as(m^2), \qquad 
 \langle\asPT\rangle_{\mu_I}  =
 {{\as(Q^2)}}  + \beta_0\frac{\as^2}{2\pi}\left(\ln\frac{Q}{\mu_I}+
    \frac{K}{\beta_0}+1\right). 
$$

\subsection{Heavy quark spectra}

 To the best of my knowledge, the first semi-rigorous attempt to put
 IR finite $\alpha_s$ at work in the pQCD context has been made in the
 pre-renormalon epoch, in the late 80s -- early 90s, in the context of
 fragmentation of heavy quarks {$H$}.

\noindent
\begin{minipage}[t]{0.4\tw}
 \includegraphics[width=\tw]{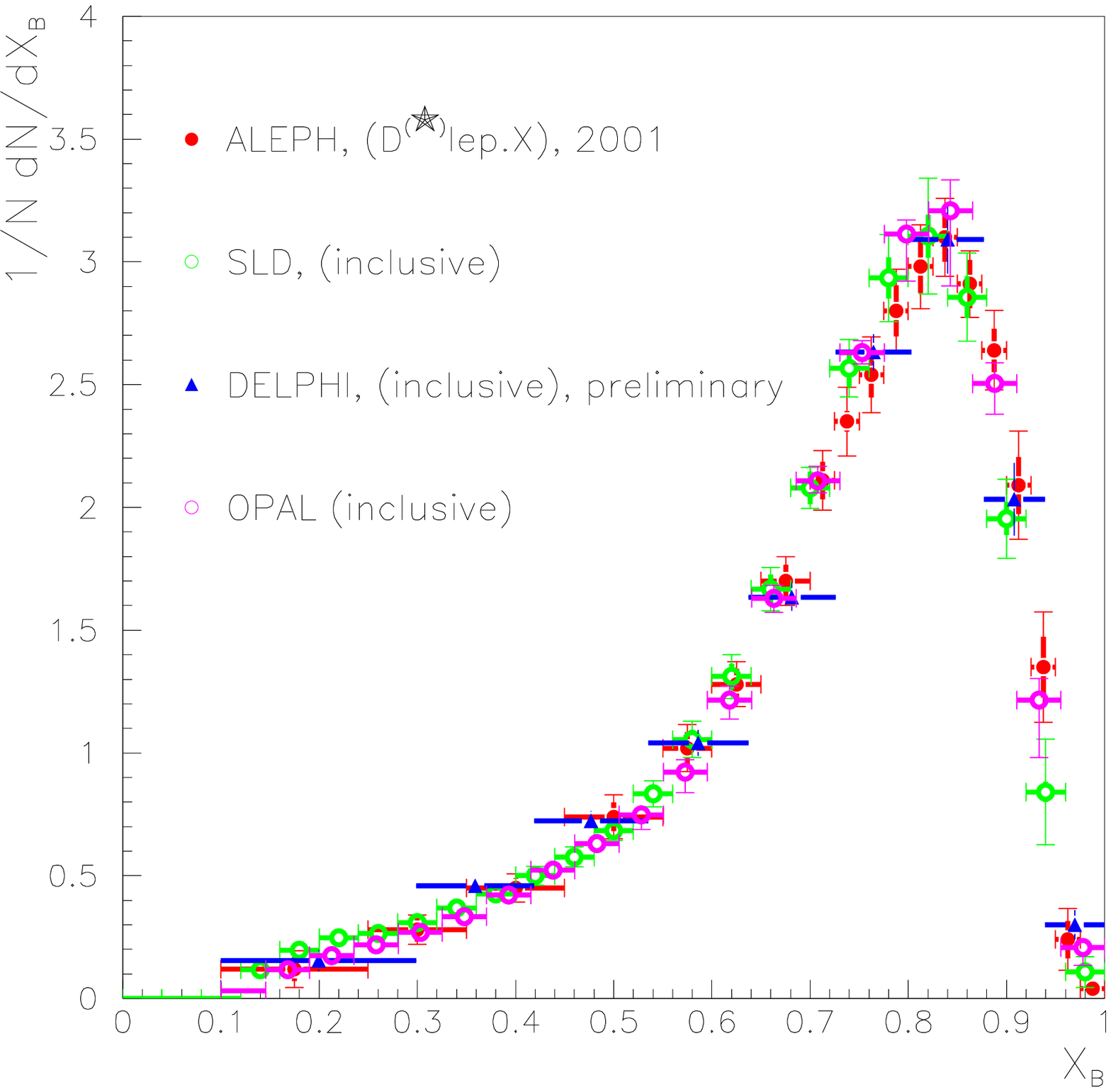}%
\end{minipage} \hfill
\begin{minipage}[b]{0.55\tw}
  Thanks to {$M_H\gg\LQCD$}, the inclusive fragmentation function $
  H\to H(x)+\ldots $ becomes collinear finite and is formally
  PT-calculable.  But only {\em formally}, since for large {$
  x$}-Feynman, {$ 1-x\la{\LQCD}/{M_H},$} one hits the NP
  domain~\cite{DKThq}.  However, it is this -- the most interesting --
  region where the fragmentation function is sitting!  (the ``leading
  heavy quark effect'').

 Fragmentation Functions (FFs), as well as space-like parton
 distributions (SFs), are
\end{minipage}\\
 known to be IR safe: for soft gluons {virtual} corrections {cancel}
 against {real} radiation contributions.  This cancellation, however,
 breaks down at the edge of the phase space, and the distributions
 acquire typical DL Sudakov suppression factors.  Here the observable
 becomes {sensitive to soft gluons} --- those very beasts that are the
 first to {enter the NP domain.}  As a result, the FF gets a NP power
 suppressed {$\cO{1/M_H}$} correction, {{\em quantifiable in terms of
 the IR coupling}}.

\subsection{A brief time-line}

 Thus, one has to compare next-to-leading {PT + NP} predictions to
 data, fitting for {$\as(Q^2)$} and {$\alpha_0$} (IR-average coupling)
 in \eqref{eq:fullNP}, in a hope to see that both $\as$ and $\alpha_0$
 will turn out to be {{\em independent of the observable}}.  
 
 The {\em power-corrections-to-event-shapes}\/ business underwent
 quite an evolution. Its dramatic element was largely due to
 impatience of experimenters who were too fast to feast on theoretical
 predictions before those could possibly reach a ``{\em well-done}\/''
 cooking status.  The new {(PT-obtainable)} coefficients {$ c_\cV$} in
 \eqref{eq:NPpar} evolved with time as shown by the following charts
 and figures elaborated by G.P.~Salam \cite{Salam:2002vj}.

\noindent
{{[95--96]}} \qquad {\bf Naive massive gluon approach}: \\
\begin{minipage}{0.57\tw}
\begin{tabular}[c]{c||c|c|c|c|c|c}%
\scriptsize    ${\cV}$ & $\tau\!=\!1\!\!-\!T$ & $C$ & $\rho$ & $\!\rho_h\!$
& $B_T$ & $B_W$\\ \hline 
${ c_\cV}$ & $2 \>(3.66)$ {?} & $3\pi$& $1$ & $2$
& $4 \ln\! Q$ & $4 \ln\! Q$
  \end{tabular}

\vspace{2mm}
 The fact that this table contains two competing predictions for the
 same variable, illustrates an intrinsic uncertainty of the then
 standard {\em naive approach}.

 {\em NB:}\/ It is fortunate that the comparison shown on the right
 insert was not available then (the $B$ and $\rho$ data appeared
 later). If it were, the whole business might have been abandoned as
 completely unsatisfactory.
\end{minipage}
\hfill
\begin{minipage}{0.4\tw}
  \epsfig{file=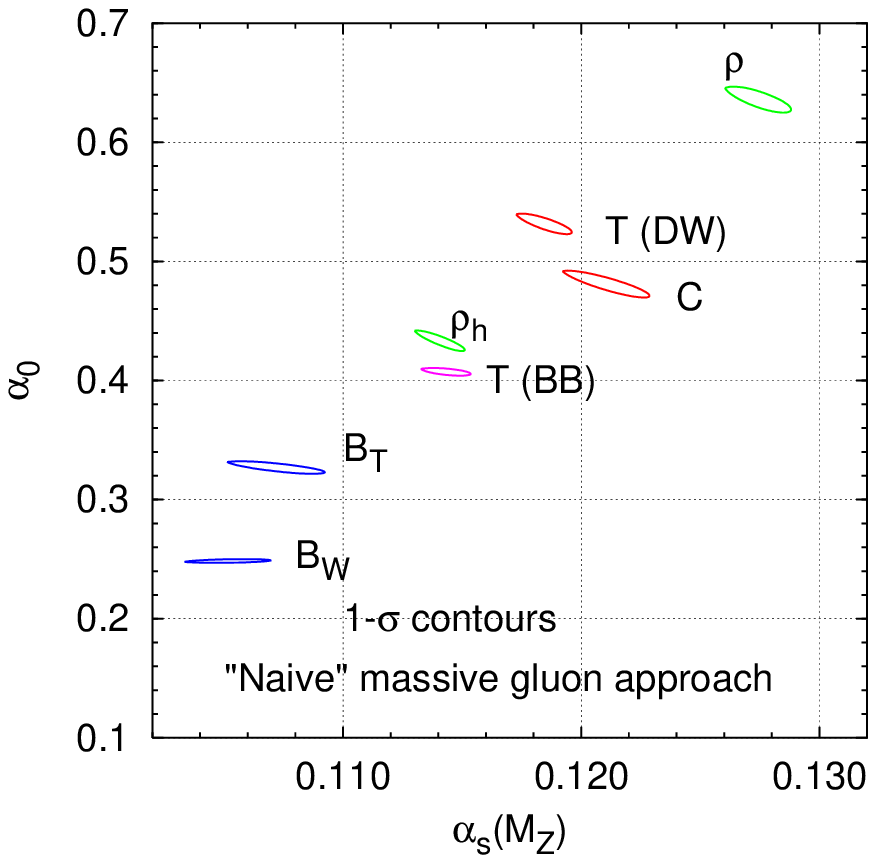,width=\textwidth}
\end{minipage}%

\noindent
{[97--00]} \qquad {\bf In-depth analysis}:\\ 
 The situation has dramatically improved when the in-depth analysis
 had been carried out which included
\smallskip 

\noindent
\begin{minipage}{0.57\tw}
\begin{itemize}%
\item understanding the PT--NP interface,
\item resummation of log-%arithmically 
 enhanced PT effects in NP contributions
\item and, finally, introduction of the so-called ``Milan factor'' 
 based on the two-loop analysis of NP terms:
\end{itemize}%
{\renewcommand{\arraystretch}{0.8}}
\begin{tabular}{c||c|c|c|c|c|c}
 $\cV$ & $\tau$ & $C$ & $\rho$ & $\!\rho_h\!$ & $B_T$ & $B_W$ \\ \hline
 ${\displaystyle {c_\cV}\!/\!{{\cal M}}}$ & ${2}$ & $3\pi$
 & $1$ & ${1}$ & {$\! \frac{\pi}{2\sqrt{C_F \as}}\!$} &
 {$\!\frac{\pi}{4\sqrt{2C_F \as}}\!$} 
\end{tabular}
\end{minipage}
\hfill
\begin{minipage}{0.4\tw}
  \epsfig{file=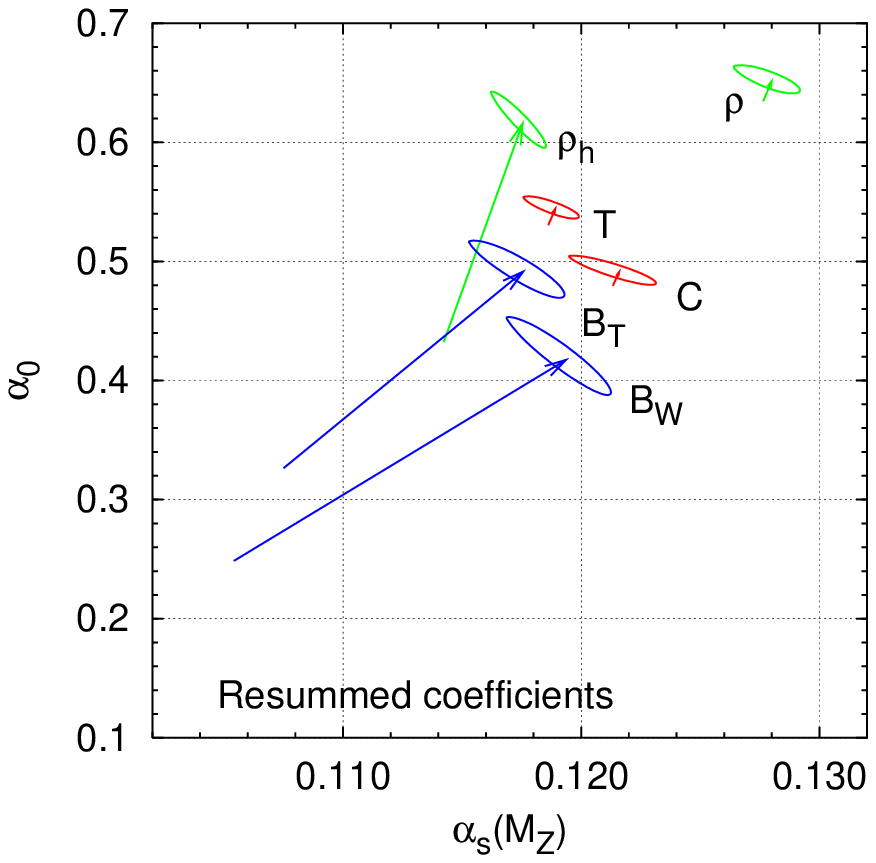,width=\textwidth}
\end{minipage}%

\noindent
\begin{minipage}{0.57\tw}
 {{[01-- ]}} \qquad {\bf Mass Effects}: \\ Finite hadron mass effects
 were recently taken care of by Salam and Wicke \cite{SW}. Their
 analysis called for special attention to be paid to the definition of
 observables at the hadron level, so as to ensure that hadron masses
 do not lead to trivial kinematic contributions that break
 universality.
\end{minipage} \hfill
\begin{minipage}{0.4\tw}
  \epsfig{file=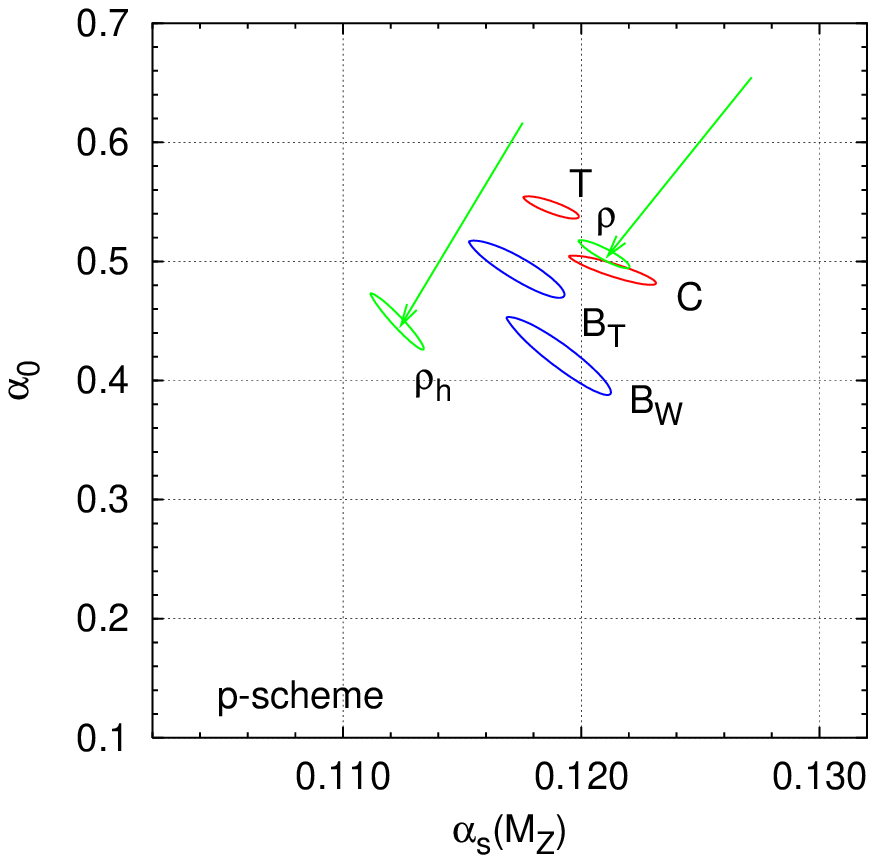,width=\textwidth}
\end{minipage}

 \subsection{Event shape distributions}

 The same technology turned out to be applicable to event shape {\em
 distributions}, {$ dN/d\cV$}.  Here the {genuine NP physics}
 manifests itself, basically, in {\em shifting}\/ the corresponding PT
 spectra, in ${ \cV}$ variable, by an amount proportional to
 {$1/Q$}~\cite{DW}.
 Distributions turned out to be an important addition to the menu.
 Firstly, to study functions is more informative and revealing than
 numbers.
 Secondly, {\em distributions}\/ are two-parameter objects
 and they don't allow essential NP contributions to be hidden under
 the carpet
 by RG-improving PT series, as the {\em means}\/ apparently
 do~\cite{Hamacher}. 

 Last but not least, it was the studies of {\em event shape
 distributions}\/ that allowed theorists to better understand what 
% the hell 
 they have been doing, thanks to pedagogical lessons theorists were
 taught by {those impatient} colleagues experimenters. In particular,
 theoretical understanding of the physics of NP effects in jet
 broadening distributions which revealed a delicate NP--PT interplay
 was triggered by the deep study of the problem pioneered by the JADE
 group~\cite{PMF}.

  NP effects aside, a purely perturbative part of the event shape
  distributions analysis is not peanuts either. All order resummation
  of logarithmically enhanced contributions was gradually carried out,
  at the next to leading order, for a vast number of observables ---
  multi-jet cross sections, thrust, $C$ parameter, jet broadenings,
  various aplanarity characteristics of three-jet events (like $D$
  parameter, $k_{\mbox{\scriptsize out}}$) etc. --- first in $e^+e^-$
  annihilation and then also in DIS, Drell--Yan processes and
  (partially) in the hadron--hadron collisions environment.

  This laborious business (with an average of $\la$ 1 observable per
  paper) turned out to be a very {\em error-prone}\/ one as well.
  Curiously, among ``professional resummers'', only about $\cO{10\%}$
%{[Trentadue, Lucenti, Smye]} 
  can say that {\em all}\/ their final results were correct to the
  accuracy claimed.  It should be noted that such a large yield of
  ``wrong papers'' (or rather of ``authors who erred'') is mostly due
  to one singular reason: a new previously overlooked class of NLLO
  corrections that hit specifically the so-called {\em non-global
  observables}\/ recently uncovered and baptized by Mrinal Dasgupta
  and Gavin Salam~\cite{NonGlob}.

  The good news is, one does not need theorists anymore anyway.  The
  {\bf C}omputer {\bf A}utomated {\bf E}xpert {\bf
  S}emi-{\bf{A}}nalytical {\bf R}esummation package has been developed
  by Andrea Banfi, Gavin Salam and Giulia Zanderighi that will analyse
  for you any observable you might fancy to invent, find out whether
  this observable is legitimate (collinear and infrared
  safe\footnote{there is a subtlety here: your observable has to be
  {\em recursively}\/ CIS, see~\cite{BSZ}. Fortunately, you have to be
  real wicked to invent one that isn't.}
  and global) and spit out the NLLO resummed
 prediction~\cite{BSZ}. Implementing the matching with exact low order
 matrix element calculations is underway~\cite{Giulia} and
 incorporating leading power suppressed NP effects is planned.

\subsection{Verifying Universality hypothesis}

\noindent
\begin{minipage}{0.45\tw}
 \includegraphics[height=\tw,width=\tw]{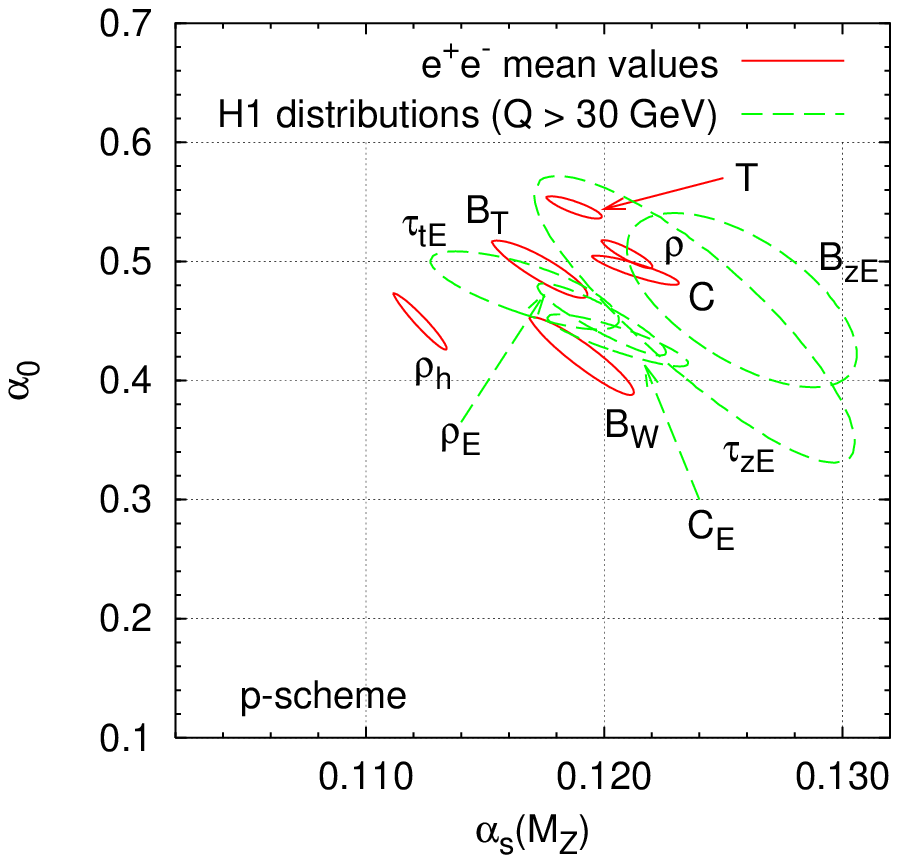}
\end{minipage}
\hfill
\begin{minipage}{0.5\tw}
 Theory \&\ Phenomenology of {$1/Q$} suppressed effects in event shape
 observables, both in $e^+e^-$ annihilation and DIS, pointed at the
 {\em average}\/ value of the infrared coupling
$$
 \displaystyle {\alpha_0} \equiv \frac1{2\,\GeV}
 \int_0^{\displaystyle 2\,\GeV} dk\>\alpha_s(k^2) \> \sim\> 0.5 \,.
$$
 As the recent analysis by Dasgupta and Salam shows~\cite{DGDIS}, the
 expected universality holds within a reasonable $15\%$ margin.
\end{minipage}

 The {Universality Hypothesis} is the key ingredient of the game: the
 new NP parameter ${\alpha_0}$ must inherit the {\em universal
 nature}\/ of the QCD coupling itself.

 Let us remark that the characteristic value of the typical PT
 expansion parameter $\as/\pi$ turns out to be numerically small,
 $\sim 0.17$ (which may open intriguing possibilities).  Moreover, it
 is comfortably above the so-called critical value of the IR coupling
 that is necessary to trigger the {\em super-critical light quark
 confinement}\/ mechanism suggested by V.~Gribov~\cite{BH} (for a
 recent review of, and around, the Gribov programme of attacking
 colour confinement see~\cite{DoKh}).

 It should be said that some recent analyses present a less positive
 picture of universality. Caution is needed in making strong claims.
 In $e^+e^-$, the majority of the analyses did not take finite
 hadron-mass effects into proper consideration; some fits look
 contradictory between different experiments (notably for $B_T$); two
 specific observables (those that select a wide/heavy hemisphere,
 $B_W$ and $\rho_h$) seem to misbehave, and probably require further
 theoretical insight; the potentially powerful technique of examining
 events with a hard final-state photon, so as to reduce the effective
 hadron cms energy and provide a lever arm in $Q$, is marred by the
 use of an incorrect assumption of factorisation of the QCD and QED
 matrix elements (this is especially of concern for mean values, since
 they are dominated by hard, non-factorising, emissions). In DIS, most
 mean values seem to lead to a highish value for $\as$, especially for
 $T_Z$ (which also has an uncomfortably low $\alpha_0$ value), for
 reasons that are not yet understood; at the same time the usually
 very tricky broadening measure seems to behave properly, even in the
 distribution.

 From the above it is clear that the universality hypothesis still
 remains to be definitively (dis)proved, with open issues both
 experimentally and theoretically. To better understand the physics of
 non-perturbative effects in jets, it is mandatory to extend the
 studies to multi-jet final-states, not only in $e^+e^-$ and DIS, but
 also, just as importantly, in hadron-hadron scattering.\footnote{The
 above overview has been produced together with G.P.~Salam.}

\section{Heavy Ions, small distances and Jets}

 Ours are the times of {\em A New Hope}. On the practice side, it is
 due to the full swing operation of the RHIC heavy ion facility at
 Brookhaven. It took off/over the CERN SPS programme which had
 already supplied us with a number of puzzles. Lead or gold, a bunch
 of intriguing phenomena have been observed in $pA$ and $AA$ high
 energy collisions.  To name but a few,
\begin{itemize}
\item
 Large-{$p_t$} pion yield gets strongly {\em suppressed}\/ in central
 collisions of heavy nuclei;
\item
 Back flowing (recoiling) jets disappear --- get {\em washed away};
\item
 Relative yields of strange mesons and baryons steadily increases with
 the number of collisions tending to ``equilibrate'' three quark
 flavours;
\item
 Leading baryons disappear from the fragmentation region
(``stopping'');
\item
 Central production of secondary baryons catches up with (if not takes
 over) that of mesons at $p_t\ga 2\,\GeV$.
\end{itemize}

 \subsection{Nucleus as ``hardener''}

 On the theory side it is becoming more and more clear that {\em small
 distances}\/ emerge naturally in the multiple scattering
 environment. Treating physics that looks a priori {\em soft}\/, such
 as inelastic diffraction off nuclei~\cite{ColDiff}, medium induced
 gluon radiation~\cite{LPM}, various phenomena that one gathers under
 the banner of the colour glass condensate (CGC) picture~\cite{CGC},
 one observes that the hardness scale that characterizes these (and
 similar) process grows invariably as
\beq\label{eq:nucl}
  Q^2\propto A^{1/3}\,.
\eeq
 A short sketch is due to illustrate this important point.

 Consider for example a non-destructive high energy hadron--nucleus
 interaction known as {\em diffraction}. To muddle through a thick
 target without causing much damage (inelastic breakup of the target),
 our projectile should interact weakly, be rather {\em transparent}.
 Since the incident hadron is a composite object, this can be achieved
 by selecting {\em compact}\/ configurations with relatively small
 separation between valence quarks in the impact parameter space,
\beq
   \sigma({\bf b}) \propto \alpha_s^2\cdot |{\bf b}|^2\,.
\eeq
 The ``transparency'' condition, 
$$ 
   \lambda \sim \frac{1}{\rho\,\sigma({\bf b})} \quad\ga\quad L, 
$$
 then gives
\beq\label{eq:lpt}
 Q^2 \> \equiv \> 1/ |{\bf b}|^2 \quad \ga \quad \as^2\rho\cdot L
\eeq
 (with $L \propto A^{1/3}$ the target thickness and $\rho$ the
 nuclear density).  A pion in such a squeezed configuration will
 fragment in the final state into two quark jets with {\em large}\/
 transverse momenta $k_\perp^2\sim Q^2$ thus {\em illuminating colour
 transparency}\/~\cite{DiffPi2Jets}.  Pion dissociation into two jets
 has been recently observed by the Fermilab E-791
 experiment~\cite{DP2Jexp} which verified jet energy, transverse
 momentum and $A$ dependences predicted by QCD.

 A very similar structure of the characteristic scale emerges in the
 problem of medium induced {\em transverse momentum broadening}\/ and
 a closely related induced {\em gluon radiation}\/ (the LPM effect,
 see below).  Here the relation mirroring \eqref{eq:lpt} is expressed
 in terms of the so-called {\em transport coefficient}
\beq\label{eq:lpmscale}
  Q^2 = \hat{q}\cdot L\,; \quad \hat{q}= \frac{\lrang{ 1/ |{\bf b}|^2
  }}{\lambda} = \frac{1}{\sigma\lambda} \int q_\perp^2\,
  d\sigma(q_\perp^2) \>\propto\> \as\rho [xG(x,Q^2)].
\eeq
 Finally, within the CGC approach to high energy phenomena in heavy
 ion interactions {\em the same}\/ characteristic parameter
 \eqref{eq:lpmscale} appears~\cite{MK} under the name of ``saturation
 scale'' $Q_s^2$.

 To conclude, the gift \eqref{eq:nucl} seems to be putting things
 under tighter pQCD control, shifting the emphasis towards underlying
 (perturbative) quark--gluon physics.  A paradoxical situation
 emerges: on the other hand, the number of puzzles is steadily
 increasing in scattering of/off nuclei; on the other hand, these
 phenomena have a good reason to be under the jurisdiction of pQCD.

 Abundant puzzles and paradoxes constitute the best imaginable setup
 for the theory as they provide potential (or rather potential
 gradient --- the field strength) for a revolutionary breakthrough.
 Jets --- the subject of this talk --- are to play a key r\^ole in
 this quest.

 \subsection{Collisions or Participants?} 

 One of the difficult questions of the physics of heavy ion collisions
 is the question of {\em scaling}. To be able to state that ``{\em
 new}\/'' physics manifests itself we better know what is to be
 expected if the physics were ``{\em old}\/''?  How to compare the
 quantity one measures in $AA$ (or $pA$) collisions with the one {\em
 simply rescaled}\/ from an elementary $pp$ interaction?  It is in
 this harmlessly looking ``simply rescaled'' that the devil resides.
  Should a given observable in $AA$ interactions scale with the number
  of {\em participating nucleons}\/ (which may be as large as $n_p\le
  2A$) or instead as the number of elementary nucleon--nucleon
  collisions ($n_c\propto n_p \cdot A^{1/3}$)?
  
 It is common wisdom to expect {\em hard}\/ interactions to scale as
 $n_c$ and {\em soft}\/ phenomena --- as $n_p$. Nothing would be
 easier than to drown, without trace, in the discussion of what is
 soft and what is hard, how hard is hard etc.  Since the purpose of
 this talk is to build up tension rather than help to release it, let
 me introduce more confusion into this (nuclear) matter.
 
 This is easy to do by recalling the QCD pattern of gluon radiation
 induced by multiple scattering of a coloured projectile in the QCD
 medium. The structure of the inclusive spectrum of medium-induced
 gluon radiation looks as follows:
\beq\label{eq:LPM}
\frac{\omega\,dn}{d\omega} \simeq {{\frac{\alpha_s}{\pi}}}\cdot
{{\displaystyle\left[\frac{L}{\lambda}\right]}} \cdot
\sqrt{\frac{\mu^2\lambda}{\omega}} ,\qquad \mu^2\lambda < \omega <
\mu^2\lambda {\left[\frac{L}{\lambda}\right]}^{2}.
\eeq
 Here $\lambda$ is the mean free path of the projectile (quark, gluon,
 \ldots), $L$ the size of the medium and $\mu$ a typical transverse
 momentum transfer in a single scattering. The first factor on the
 r.h.s.\ of \eqref{eq:LPM} corresponds to the Bethe--Heitler (BH)
 regime of independent emission of a gluon off each scattering
 centre. Taken together with the second factor (the total number of
 elementary collisions of the projectile, $n_c=L/\lambda$) this would
 correspond to secondary gluon production according to the
 $n_c$--scaling prescription. However, the third factor modifies the
 BH prediction by suppressing the yield of more energetic gluons. It
 is only the softest radiation with finite energies $\omega\sim
 \mu^2\lambda$ that follows the BH pattern.

\smallskip
\noindent
\begin{minipage}{0.55\tw}%
 This is what is going on, from the QCD point of view, in $hA$
 scattering.  Quantum mechanical coherence suppresses production of
 more energetic gluons whose yield gradually turns into that of {\em
 participant}\/ scaling ($n_p=1$ in our example). As shown by the
 accompanying schematic picture of the particle yield in $hA$
 collisions, the transition is smooth and occu-
\end{minipage}
\begin{minipage}{0.43\tw}
\resizebox{!}{0.6\textwidth}{\begin{picture}(0,0)%
\includegraphics{lpm.pstex}%
\end{picture}%
\setlength{\unitlength}{3947sp}%
\begingroup\makeatletter\ifx\SetFigFont\undefined%
\gdef\SetFigFont#1#2#3#4#5{%
  \reset@font\fontsize{#1}{#2pt}%
  \fontfamily{#3}\fontseries{#4}\fontshape{#5}%
  \selectfont}%
\fi\endgroup%
\begin{picture}(3372,2494)(-949,-3883)
\put(-449,-1711){\makebox(0,0)[lb]{\smash{{\SetFigFont{20}{24.0}{\familydefault}{\mddefault}{\updefault}{$n_c$}%
}}}}
\put(751,-2161){\makebox(0,0)[lb]{\smash{{\SetFigFont{20}{24.0}{\familydefault}{\mddefault}{\updefault}{$e^{-\eta/2}$}%
}}}}
%\put(151,-3811){\makebox(0,0)[lb]{\smash{{\SetFigFont{14}{16.8}{\familydefault}{\mddefault}{\updefault}{$y_A$}%
%}}}}
\put(2851,-3811){\makebox(0,0)[lb]{\smash{{\SetFigFont{20}{24.0}{\familydefault}{\mddefault}{\updefault}{$\eta$}%
}}}}
\put(-449,-3211){\makebox(0,0)[lb]{\smash{{\SetFigFont{20}{24.0}{\familydefault}{\mddefault}{\updefault}{${n_p}$}%
}}}}
\put(901,-3811){\makebox(0,0)[lb]{\smash{{\SetFigFont{14}{16.8}{\familydefault}{\mddefault}{\updefault}{$2\ln n_c$}%
}}}}
\end{picture}%

}
\end{minipage}\\[1mm]
% 
%\noindent
 pies a finite range in $\eta=\ln\omega$ from the fragmentation region
 of the nucleus, $\Delta\eta \simeq 2\ln n_c$. This coherent
 suppression is similar in nature to that known under the name of the
 Landau--Pomeranchuk--Migdal (LPM) effect in QED.

 The essence of the QCD LPM physics is easy to grasp. A number of
 scattering centres ($N_{\mbox{\scriptsize coh.}}$) that fall {\em
 inside the formation length}\/ of the gluon act {\em coherently}\/ as
 a single scatterer. At the same time, the gluon is subject to
 Brownian motion in the transverse momentum plane, so that
\beq 
 k_\perp^2 \simeq N_{\mbox{\scriptsize coh.}} \cdot \mu^2\,, \qquad
 N_{\mbox{\scriptsize coh.}} \simeq \frac{\ell_{\mbox{\scriptsize
 coh.}}}{\lambda} \simeq \frac1{\lambda}\cdot \frac{\omega}{k_\perp^2}. 
\eeq
Combining the two estimates results in
\beq\label{eq:kp}
  N_{\mbox{\scriptsize coh.}} \simeq
\sqrt{\frac{\omega}{\mu^2\lambda}} \qquad \mbox{and} \quad
 k_\perp^2\simeq  \sqrt{\frac{\mu^2}{\lambda}\cdot\omega}\,.
\eeq
It is the factor $ N_{\mbox{\scriptsize coh.}}^{-1}$ that describes
the coherent LPM suppression in \eqref{eq:LPM}.

 Now comes the confusion part. From \eqref{eq:kp} we observe that {\em
 more energetic}\/ gluons have typically {\em larger transverse
 momenta}. This means, in turn, that the accompanying radiation
 corresponding to {\em larger hardness scales}\/ (gluons with larger
 $k_\perp$) follows the {\em participant}\/ scaling, while the {\em less
 hard}\/ radiation (smaller $k_\perp$ and energies) obeys the {\em
 collisional}\/ scaling pattern, in striking contradiction with the
 original expectation.  It is quantum mechanics that is to be blamed for such
 a miserable failure of ``common wisdom''.

 \subsection{Colour in multiple scattering}

 The situation appears even more confusing if we recall the r\^ole of
 colour in multiple hadroproduction and try to reconcile the
 underlying colour dynamics with the high energy hadron interaction
 phenomenology based on the multi-Pomeron scattering picture.
 
 \subsubsection{Single scattering, Pomeron and accompanying gluons. }

\noindent 
\begin{minipage}{0.55\tw}
 From the QCD point of view, scattering of a pion due to one gluon
 exchange breaks coherence of the valence $q\bar{q}$ system and
 results in multiparticle production (inelastic process) via
 appearance of the two ``quark chains'' in the final state.
 These two Kogut--Susskind chains are nothing but an image of the
 Feynman plateau --- 
\end{minipage}
\hfill
\begin{minipage}{0.4\tw}
 \resizebox{!}{0.6\textwidth}{\begin{picture}(0,0)%
\includegraphics{glue_pi2.pstex}%
\end{picture}%
\setlength{\unitlength}{3947sp}%
\begingroup\makeatletter\ifx\SetFigFont\undefined%
\gdef\SetFigFont#1#2#3#4#5{%
  \reset@font\fontsize{#1}{#2pt}%
  \fontfamily{#3}\fontseries{#4}\fontshape{#5}%
  \selectfont}%
\fi\endgroup%
\begin{picture}(6084,3500)(253,-5965)
\end{picture}%
}%
\end{minipage}
\smallskip

\noindent
 the cut Pomeron.  The same is true for inelastic scattering of a {\em
 proton}. Since one gluon exchange leaves a {\em diquark}\/ unbroken,
 the same two chains develop, the only difference being that one of
 the chains starts off from a valence diquark and therefore gives
 birth, as a rule, to a {\em leading baryon}\/ in the final state.

 This two-quark-chain imagery of an universal Pomeron seems to be in a
 perfect accord with the LPHD view according to which secondary
 hadrons in the final state originate from accompanying gluon
 radiation. Indeed, consider gluon radiation (momentum $k$, colour
 $a$) in course of scattering of a projectile off the gluon field
 (momentum transfer $q$, colour $b$):

 \resizebox{!}{0.25\textwidth}{\begin{picture}(0,0)%
\includegraphics{coher2.pstex}%
\end{picture}%
\setlength{\unitlength}{3947sp}%
\begingroup\makeatletter\ifx\SetFigFont\undefined%
\gdef\SetFigFont#1#2#3#4#5{%
  \reset@font\fontsize{#1}{#2pt}%
  \fontfamily{#3}\fontseries{#4}\fontshape{#5}%
  \selectfont}%
\fi\endgroup%
\begin{picture}(8377,2820)(-47,-2011)
\put(4322,-49){\makebox(0,0)[rb]{\smash{{\SetFigFont{20}{24.0}{\rmdefault}{\bfdefault}{\updefault}{T}%
}}}}
\put(4492,121){\makebox(0,0)[rb]{\smash{{\SetFigFont{20}{24.0}{\rmdefault}{\bfdefault}{\updefault}{b}%
}}}}
\put(5382,-59){\makebox(0,0)[rb]{\smash{{\SetFigFont{20}{24.0}{\rmdefault}{\bfdefault}{\updefault}{T}%
}}}}
\put(5542,131){\makebox(0,0)[rb]{\smash{{\SetFigFont{20}{24.0}{\rmdefault}{\bfdefault}{\updefault}{a}%
}}}}
\put(1882,-39){\makebox(0,0)[rb]{\smash{{\SetFigFont{20}{24.0}{\rmdefault}{\bfdefault}{\updefault}{T}%
}}}}
\put(2052,131){\makebox(0,0)[rb]{\smash{{\SetFigFont{20}{24.0}{\rmdefault}{\bfdefault}{\updefault}{b}%
}}}}
\put(692,-59){\makebox(0,0)[rb]{\smash{{\SetFigFont{20}{24.0}{\rmdefault}{\bfdefault}{\updefault}{T}%
}}}}
\put(852,131){\makebox(0,0)[rb]{\smash{{\SetFigFont{20}{24.0}{\rmdefault}{\bfdefault}{\updefault}{a}%
}}}}
\put(7282,-59){\makebox(0,0)[rb]{\smash{{\SetFigFont{20}{24.0}{\rmdefault}{\bfdefault}{\updefault}{T}%
}}}}
\put(7442,131){\makebox(0,0)[rb]{\smash{{\SetFigFont{20}{24.0}{\rmdefault}{\bfdefault}{\updefault}{c}%
}}}}
\put(3076,-436){\makebox(0,0)[rb]{\smash{{\SetFigFont{29}{34.8}{\rmdefault}{\bfdefault}{\updefault}{+}%
}}}}
\put(1351,-1261){\makebox(0,0)[rb]{\smash{{\SetFigFont{29}{34.8}{\familydefault}{\mddefault}{\updefault}{q}%
}}}}
\put(751,614){\makebox(0,0)[rb]{\smash{{\SetFigFont{20}{24.0}{\rmdefault}{\bfdefault}{\updefault}{a}%
}}}}
\put(2026,-2011){\makebox(0,0)[rb]{\smash{{\SetFigFont{20}{24.0}{\rmdefault}{\bfdefault}{\updefault}{b}%
}}}}
\put(6086,-426){\makebox(0,0)[rb]{\smash{{\SetFigFont{29}{34.8}{\rmdefault}{\bfdefault}{\updefault}{+}%
}}}}
\put(1426,464){\makebox(0,0)[rb]{\smash{{\SetFigFont{29}{34.8}{\familydefault}{\mddefault}{\updefault}{k}%
}}}}
\put(7012,-1219){\makebox(0,0)[rb]{\smash{{\SetFigFont{17}{20.4}{\rmdefault}{\bfdefault}{\updefault}{abc}%
}}}}
\put(6532,-1079){\makebox(0,0)[rb]{\smash{{\SetFigFont{25}{30.0}{\rmdefault}{\bfdefault}{\updefault}{if}%
}}}}
\put(7596,-1971){\makebox(0,0)[rb]{\smash{{\SetFigFont{20}{24.0}{\rmdefault}{\bfdefault}{\updefault}{b}%
}}}}
\put(8261,-1276){\makebox(0,0)[rb]{\smash{{\SetFigFont{20}{24.0}{\rmdefault}{\bfdefault}{\updefault}{a}%
}}}}
\end{picture}%
}

\noindent
The sum of the three relevant amplitudes reduces to
 \[
    -\frac{{\bf k}_\perp}{{\bf k}_\perp^2}{{
    T^bT^a}} + \frac{{\bf k}_\perp}{{\bf
    k}_\perp^2}{{T^aT^b}} + \frac{{\bf
    q}_\perp-{\bf k}_\perp}{({\bf q}_\perp-{\bf
    k}_\perp)^2}{if_{abc}{ T^c}}
 = if_{abc}{ T^c}\cdot \left[\frac{{\bf k}_\perp}{{\bf
    k}_\perp^2}+  \frac{{\bf
    q}_\perp-{\bf k}_\perp}{({\bf q}_\perp-{\bf
    k}_\perp)^2}\right].
\]
 Since the colour group commutation relation $[T^aT^b]=if_{abc}T^c$ is
 the same for any projectile (an arbitrary colour object with
 generator $T$), the accompanying radiation intensity turns out to be
 universal and proportional to the ``colour charge'' of the
 $t$-channel exchange, $(if_{abc})^2\propto N_c={\bf 3}$.  This
 universality is a direct consequence of conservation of the {\em
 colour current}.

 \subsubsection{Multiple scattering, AFS and ``colour capacity''}

 Problems start emerging when we turn to multiple scattering. Let us
 exchange two gluons and calculate the average {\em colour charge}\/
 of the $t$-channel two-gluon system (with bold numbers standing for
 colour factors):
\[
  \frac{{ 1}}{64}\cdot{\bf 0} + \frac{{ 8}+{ 8}}{64}\cdot{\bf 3} +
  \frac{{10}+{\overline{10}}}{64}\cdot{\bf 6} + \frac{{
  27}}{64}\cdot{\bf 8} = {\bf 6} = 2\cdot{\bf 3}.
\]
 This looks satisfactory since the doubling of the radiation intensity
 will translate into the double density of produced hadrons (two cut
 Pomerons). However, validity of this result is questionable since in
 the above calculation we have completely ignored the nature of the
 projectile. Indeed, if we take a single quark or a colour-neutral
 $q\bar{q}$ system (meson) as a projectile, the transition current can
 only be a colour octet (with a bit of a singlet), whatever the number
 of exchanged gluons! This means that multiple scattering of a
 ($q\bar{q}$) pion will never produce anything but a single-Pomeron
 particle density (unless we look specifically close to the {\em
 target}\/ fragmentation, cf.\ the LPM discussion above).

 In spite of the proton looking more {\em capacious}\/ a projectile,
 the two-Pomeron exchange cannot be realized on a {\em
 valence-built}\/ proton either. Indeed, since a 3-quark system can be
 repainted only into colour representations ${\bf 1}$ + ${\bf 8}$ +
 ${\bf 8}$ + ${\bf 10}$ (altogether 27 states), the average density of
 the gluon accompaniment will be
\[
  \frac{{1}}{27}\cdot{\bf 0} + \frac{{8}+{ 8}}{27}\cdot{\bf 3} +
  \frac{{10}}{27}\cdot{\bf 6} = {\bf 4} = 1.5\times \>\mbox{Pomeron
  (?!)}
\]
 Let us remark that we should have expected this weird ``1.5 Pomeron''
 yield from the naive chain consideration already, since it is {\em
 three}\/ quark chains that develop in course of independent
 fragmentation of the valence quarks of the broken proton\footnote{By
 the way, the baryon number of a broken proton naturally sinks into
 the sea by 2--3 rapidity units (stopping).}  (while one would need
 {\em four}\/ to picture two cut Pomerons).

 QCD scenario of hadroproduction being due to coherent gluon
 accompaniment seems to {\em invalidate}\/ the multi-Pomeron exchange
 picture. This striking statement is, however, as true as it is not
 new.  Recall the good old Amati-Fubini-Stanghellini puzzle.

 Successive scatterings of a single parton \underline{do not produce}
 {\em branch points}\/

\noindent
\begin{minipage}{0.5\tw}
 in the complex angular momentum plane (Reggeon loops).  Instead, it
 is the {\em Mandelstam construction}\/ that generates ``Reggeon
 cuts'', with Pomerons attached not to one but to separate ---
 {\em coexisting}\/ --- partons.
\end{minipage} \hfill
\begin{minipage}{0.45\tw}
\resizebox{!}{0.35\textwidth}{\begin{picture}(0,0)%
\includegraphics{mandel.pstex}%
\end{picture}%
\setlength{\unitlength}{3947sp}%
\begingroup\makeatletter\ifx\SetFigFont\undefined%
\gdef\SetFigFont#1#2#3#4#5{%
  \reset@font\fontsize{#1}{#2pt}%
  \fontfamily{#3}\fontseries{#4}\fontshape{#5}%
  \selectfont}%
\fi\endgroup%
\begin{picture}(4850,1699)(-321,-3108)
\put(1426,-2236){\makebox(0,0)[lb]{\smash{{\SetFigFont{14}{16.8}{\familydefault}{\mddefault}{\updefault}{=0}%
}}}}
\end{picture}%
}
\end{minipage}
\smallskip

\noindent
 Thus, to have {$n_c$}-gluon exchange produce {(up to)} {$n_c$} times
 enhanced density of the hadron plateau ($n_c$ cut Pomerons), one must be
 able to find {$n_c$} {\em independent}\/ {(incoherent)} {\em
 partons}\/ inside the projectile.

 The answer to whether the final hadron yield follows the collision or
 participant scaling depends on the question (as it does so often in
 quantum mechanics). It depends on what are we looking at and becomes
 the question of {\em resolution}.

 To be able to absorb $n_c$ gluons {\em incoherently}\/ in order to
 give rise to $n_c$ Pomeron chains (collision scaling), our projectile
 has to have sufficient ``colour capacity''. We must compare the
 number of collisions {$n_c$} with the number of {\em resolved
 partons}\/ inside the projectile,
\[
  {C}(x_h,Q_{\mbox{\scriptsize res}}) \>\simeq\>
  \int_{x_h}^{x_{\mbox{\scriptsize proj}}} \frac{dx}{x}\,
  \left[xG_{\mbox{\scriptsize proj}}(x,Q_{\mbox{\scriptsize
  res}}^2)\right] \>+\> \mbox{[valence stuff]}.
\]
 By increasing $Q_{\mbox{\scriptsize res}}$ (transverse momentum of
 the registered hadrons $h$) and/or by moving further away from the
 projectile in rapidity (increasing $\ln(x_{\mbox{\scriptsize
 proj}}/x_h)$) we will gradually get rid of colour coherence ($n_p$
 scaling) and approach the $n_c$ scaling.

 \subsection{Confinement in new environment} 

 In the framework of the standard multi-Pomeron picture\footnote{e.g.,
 in the successful Dual Parton Model of Capella, Kaidalov et
 al.~\cite{CK}} one includes {\em final state interactions}\/ to
 explain spectacular heavy ion phenomena like {$J/\psi$} suppression,
 enhancement of {strangeness} production and alike.  ``Final state
 interaction'' is a synonym to ``non-independent fragmentation'' (one
 hears about cross-talking Pomerons, overlapping strings, ``string
 ropes'', \ldots, you name it).

  From the point of view of the {\em colour dynamics}\/, in {$pA$} and
 {$AA$} environments we face an intrinsically new, unexplored,
 question: After the pancakes separate, at each impact parameter we
 have the colour field strength corresponding to
 {$n_p/\mbox{fm}^2\propto A^{1/3}$} ``strings''.  {How does the vacuum
 break up} in such a -- {\em stronger than usual}\/ -- colour field?
 Imagine we stretch a high density field (pull apart over-charged
 capacitor plates). Will it go like

\smallskip
\noindent
\begin{minipage}{0.4\tw}
\begin{center}
 \includegraphics[width=0.95\tw]{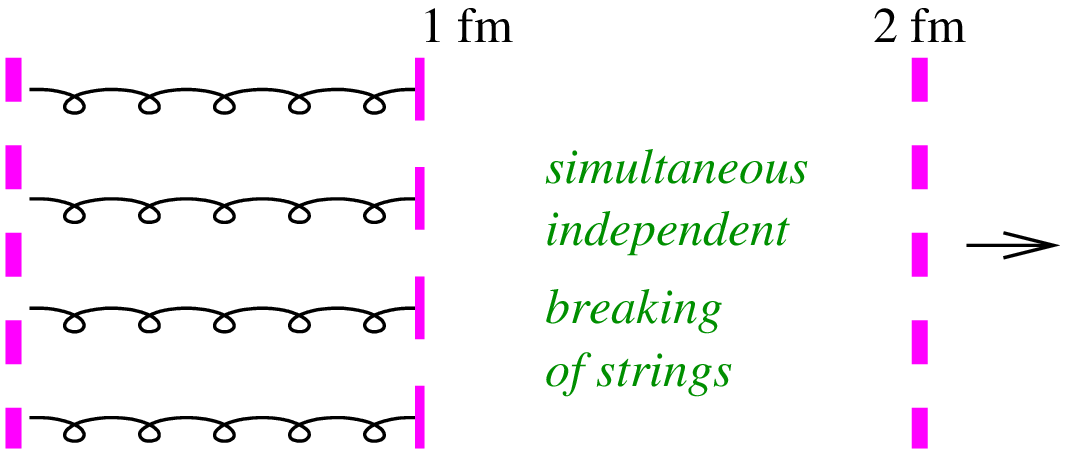}\\
 {\bf BOOOOM}
\end{center}
\end{minipage}
\quad or rather \qquad
\begin{minipage}{0.4\tw}
\begin{center}
 \includegraphics[width=0.95\tw]{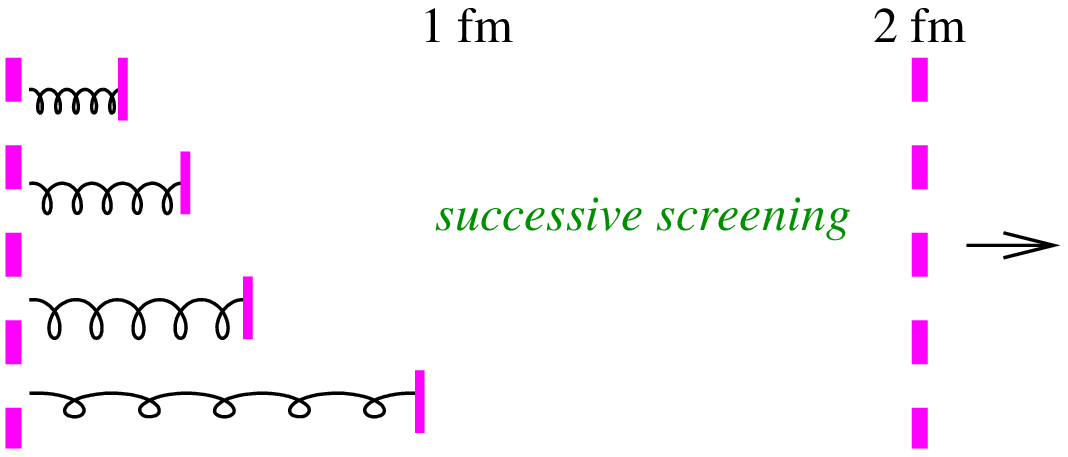}\\
 {\bf TA-TA--TA---TA} ?
\end{center}
\end{minipage}

\vspace{2mm}
\noindent 
 The first scenario corresponds to four cut Pomerons (quadruple
 multiplicity but standard particle abundances). In the second one
 vacuum break-up occurs at {\em smaller distances}\/ and therefore
 will provide, in particular, a free {\em strangeness}\/ lunch
 (together with other sweet cookies).

\section{Conclusions}

 Jets as a PT instrument did the job they've been asked to perform: to
 verify the nature of the fundamental fields of the underlying QFT by
 measuring quark and gluon {spins}, to establish {$SU_c(3)$} as the
 true QCD gauge group.

 Jets did more than that: studying inclusive energy spectra of
 {(relatively soft)} hadrons {\em inside}\/ jets, {and} soft hadron
 multiplicity flows {\em in-between}\/ jets taught us an important
 lesson, {or rather have sent us a hint,} about the non-violent nature
 of hadronization {(``{\em soft confinement}\/'')}.

 On the nuclear side, Jets stemming from diffractive hadron
 dissociation on nuclei reveal the internal small-distance structure
 of hadron projectiles; Jets that are produced in, and muddle through,
 the colour soup left behind head-on collisions of heavy nuclei bear
 information about new peculiar QCD media (``{\em jet quenching}\/'').

 Jets --- their appearance, disappearance, internal structure
 (particle abundances, shape observables, angular profiles, etc.) ---
 are steadily becoming a Non-Perturbative tool for elucidating the
 physics of hadronization.

%==================================================

\end{document}